\documentclass[fleqn,usenatbib]{mnras}

\usepackage{newtxtext,newtxmath}

\usepackage[T1]{fontenc}

\usepackage{graphicx}	
\usepackage{amsmath}	

\DeclareRobustCommand{\VAN}[3]{#2}
\let\VANthebibliography\thebibliography
\def\thebibliography{\DeclareRobustCommand{\VAN}[3]{##3}\VANthebibliography}

\usepackage{amsfonts}
\usepackage{latexsym}
\usepackage{color,xcolor}
\usepackage{hyperref}
\usepackage{verbatim}
\usepackage{slashed}
\usepackage{notes2bib}
\usepackage{tikz}

\newcommand{\bea}{\begin{eqnarray}}
\newcommand{\eea}{\end{eqnarray}}

\newcommand{\GeV}{\,{\rm GeV}}
\newcommand{\MeV}{\,{\rm MeV}}
\newcommand{\keV}{\,{\rm keV}}
\newcommand{\eV}{\,{\rm eV}}

\newcommand{\cm}{\,{\rm cm}}

\newcommand{\lya}{Ly$\alpha$ }
\newcommand{\Hy}{{\rm H}}
\newcommand{\alf}{\textrm{Alfv\'en }}

\def\bfk{{\bf k}}
\def\bfp{{\bf p}}

\title[Primordial black holes and the interstellar medium]{A constraint on light primordial black holes \\from the interstellar medium temperature}

\author[Hyungjin Kim]{Hyungjin Kim$^1$\\
$^1$Department of Particle Physics and Astrophysics, Weizmann Institute of Science, Rehovot 7610001, Israel
}

\author[Hyungjin Kim]{
Hyungjin Kim,$^{1,2}$\thanks{E-mail: hyungjin.kim@desy.de}
\\ 
$^{1}$Department of Particle Physics and Astrophysics, Weizmann Institute of Science, Rehovot 7610001, Israel
\\
$^{2}$DESY, Notkestrasse 85, 22607 Hamburg, Germany}

\date{}

\pubyear{2021}

\begin{document}
\label{firstpage}
\pagerange{\pageref{firstpage}--\pageref{lastpage}}
\maketitle

\begin{abstract}
Primordial black holes are a viable dark matter candidate. They decay via Hawking evaporation. Energetic particles from the Hawking radiation interact with interstellar gas, depositing their energy as heat and ionization.
For a sufficiently high Hawking temperature, fast electrons produced by black holes deposit a substantial fraction of energy as heat through the Coulomb interaction. Using the dwarf galaxy Leo T, we place an upper bound on the fraction of primordial black hole dark matter. For $M <  5 \times 10^{-17}M_\odot$, our bound is competitive with or stronger than other bounds.
\end{abstract}


\section{Introduction}
Primordial black holes (PBHs) are a viable dark matter candidate. It is minimal in the sense that it does not require new particle state to be added to the spectrum of the standard model. 
Although it is a compelling candidate, it can only account for all of the dark matter in the mass range of PBHs $10^{-16} \lesssim M / M_\odot \lesssim 10^{-11} $. See the recent review \citep{Carr:2020gox} for details. 

PBHs are decaying dark matter. They evaporate through Hawking radiation \citep{Hawking:1974rv, Hawking:1974sw}. Classical black holes can be described as a thermodynamical system, and the spectrum of emitted particles resembles the blackbody spectrum of temperature $T = 1/8\pi G M$ with a modification due to gray body factors. Various astrophysical constraints for $M \lesssim 10^{-16}M_\odot$ exist, based on Hawking evaporation of PBHs. This includes constraints from extragalactic $\gamma$-ray flux \citep{Carr:2009jm, Arbey:2019vqx, Ballesteros:2019exr}, Galactic $\gamma$-ray flux \citep{Carr:2016hva, DeRocco:2019fjq, Laha:2019ssq, Dasgupta:2019cae, Laha:2020ivk, 2020MNRAS.tmp.2094C}, cosmic microwave background \citep{Clark:2016nst, Stocker:2018avm, Poulter:2019ooo, Acharya:2019xla, Acharya:2020jbv}, cosmic rays \citep{Boudaud:2018hqb}, and the 21cm absorption \citep{Clark:2018ghm}.

\begin{figure}
\centering
\includegraphics[width=\columnwidth]{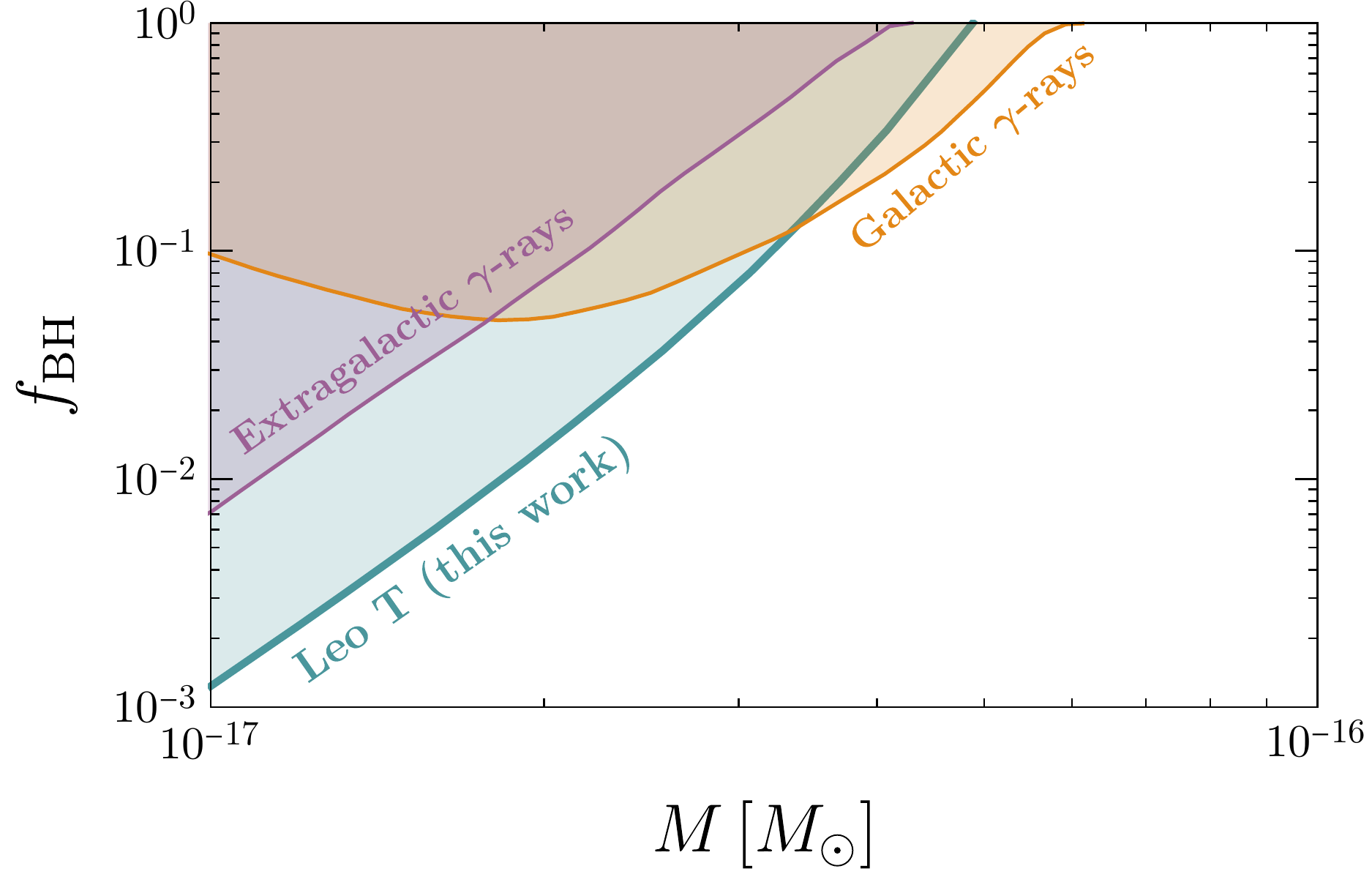}
\caption{Constraints on the PBH dark matter fraction $f_{\rm BH}$ is shown. Our result is shown as a turquoise shaded region. We also show other constraints from different astrophysical considerations: Galactic $\gamma$-ray flux measured by INTEGRAL satellite \citep[orange shaded,][]{Laha:2020ivk}, and extragalactic $\gamma$-ray flux without AGN background \citep[purple shaded,][]{Ballesteros:2019exr} for comparison. 
The bound scales as $f_{\rm BH} \propto M^3$ as long as $M\gtrsim 10^{-18} M_\odot$. See the main text for details.  }
\label{fig:result}
\end{figure}

In this paper, we investigate interactions of energetic particles from Hawking radiation in the interstellar medium (ISM). We focus on $M \lesssim 10^{-16}M_\odot$. The corresponding Hawking temperature is $T_{\rm BH} \gtrsim 50\keV$, and energetic electrons can be easily produced from Hawking evaporation. Once produced, such fast electrons quickly lose most of their energy either through elastic scatterings with charged particles or through ionizing neutral atoms in the gas. A significant fraction of energy is deposited as heat, and this could contribute to ISM heating as additional source.
By considering observed properties of warm neutral medium (WNM) in the dwarf galaxy Leo T, we place a new constraint on the fraction of PBH dark matter for $M \lesssim 5\times 10^{-17} M_\odot$. Our main result is shown in Figure \ref{fig:result}.

We focus solely on the dwarf galaxy Leo T in this work. 
Leo T is particularly interesting when it comes to constraining additional heating from PHBs since (i) it is dark matter dominated, (ii) it hosts a large amount of neutral hydrogen gas, and (iii) the metal abundance is low. All these features makes Leo T particularly susceptible to additional heating from PBHs.

We note that Leo T as well as cold gas clouds in Milky Way have been already investigated to constrain scattering cross sections of WIMP dark matter with the standard model particles \citep{Bhoonah:2018wmw, Bhoonah:2018gjb, Farrar:2019qrv, Wadekar:2019xnf}. In addition, \citet{Lu:2020bmd} has recently shown that accreted disk around solar mass PBHs as well as dynamical friction could also heat interstellar gases, and that Leo T can be used to constrain PBHs for the masses above $M_\odot$.

The paper is organized as follows. In Section \ref{sec:LeoT}, we summarize the observed properties of the dwarf galaxy Leo T. We discuss the cooling and heating of ISM in Section \ref{sec:cooling_heating}, and apply them to Leo T to derive a constraint on the PBH dark matter fraction in Section \ref{sec:result}.
We discuss assumptions made in the analysis in Section \ref{sec:discussion}.

\section{Leo T}\label{sec:LeoT}
As we have mentioned in the introduction, we only consider the dwarf galaxy Leo T in this paper. It stands out among the other faint dwarf galaxies as it is dark matter dominated, and, at the same time, is gas-rich system. It is located at a distance of $420$ kpc, and hosts both cold and warm neutral gases. There is no indication of bulk rotation. The velocity dispersion is $\sigma =6.9$ km/sec, which corresponds to $T_{\rm wnm} =6000$ K. The total mass of atomic hydrogen, $M_{\rm wnm} = 2.8\times 10^5 M_\odot$, is extended over $r_{\rm wnm} =350$ pc \citep{RyanWeber:2007fb, 2018A&A...612A..26A}. From the metallicity measurement of individual stars, iron abundance in Leo T is found as $[{\rm Fe/H}] = -1.74$ \citep{2013ApJ...779..102K}.\footnote{$[{\rm X/Y}] \equiv \log_{10} (n_X/n_Y) - \log_{10}(n_X/n_Y)_\odot$ where $n_{\rm X,Y}$ is the number density of species X and Y, and the subscript $\odot$ represents the solar abundance. }

\begin{figure}
\centering
~~~\includegraphics[width=.985\columnwidth]{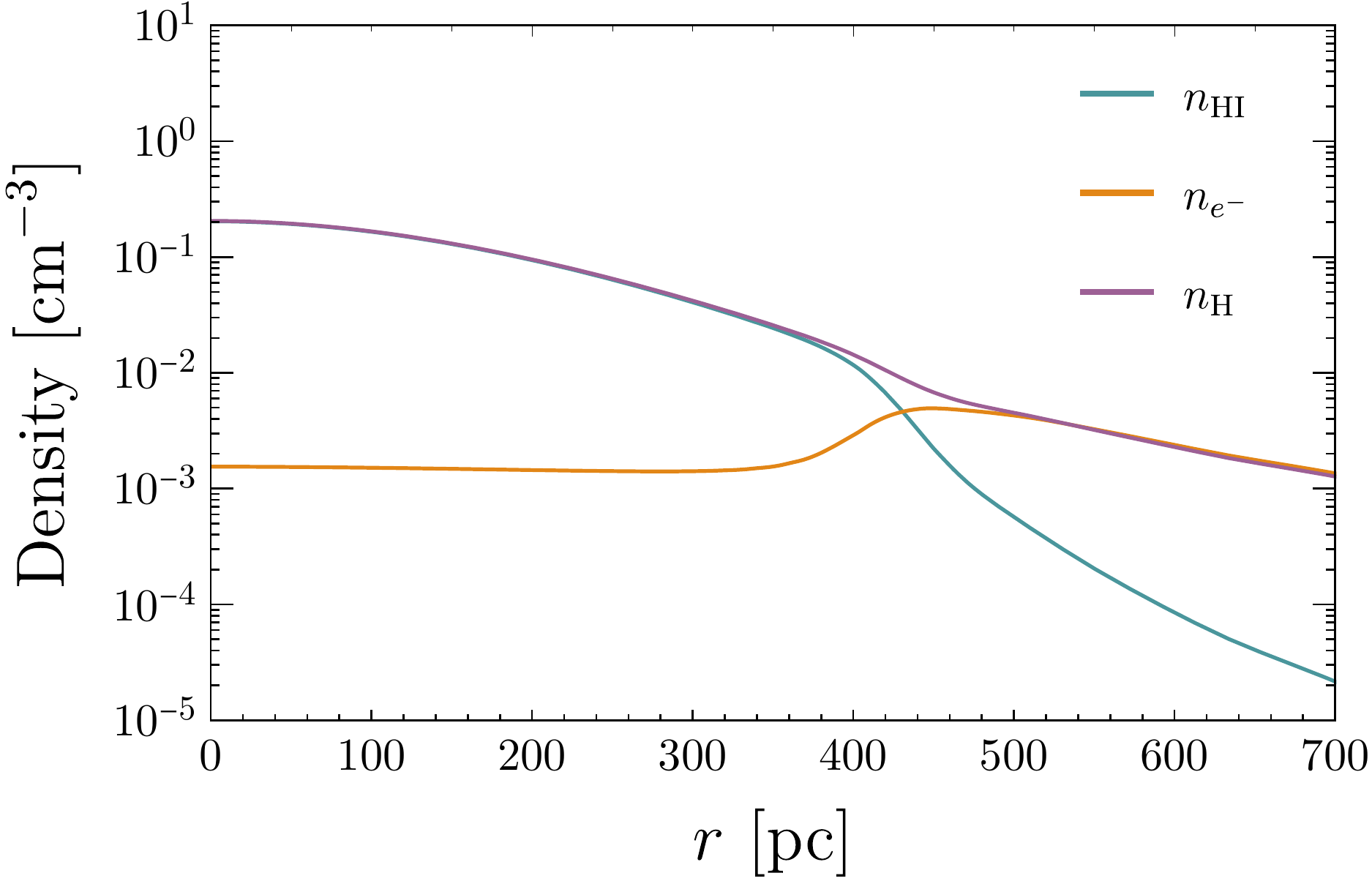}
\includegraphics[width=\columnwidth]{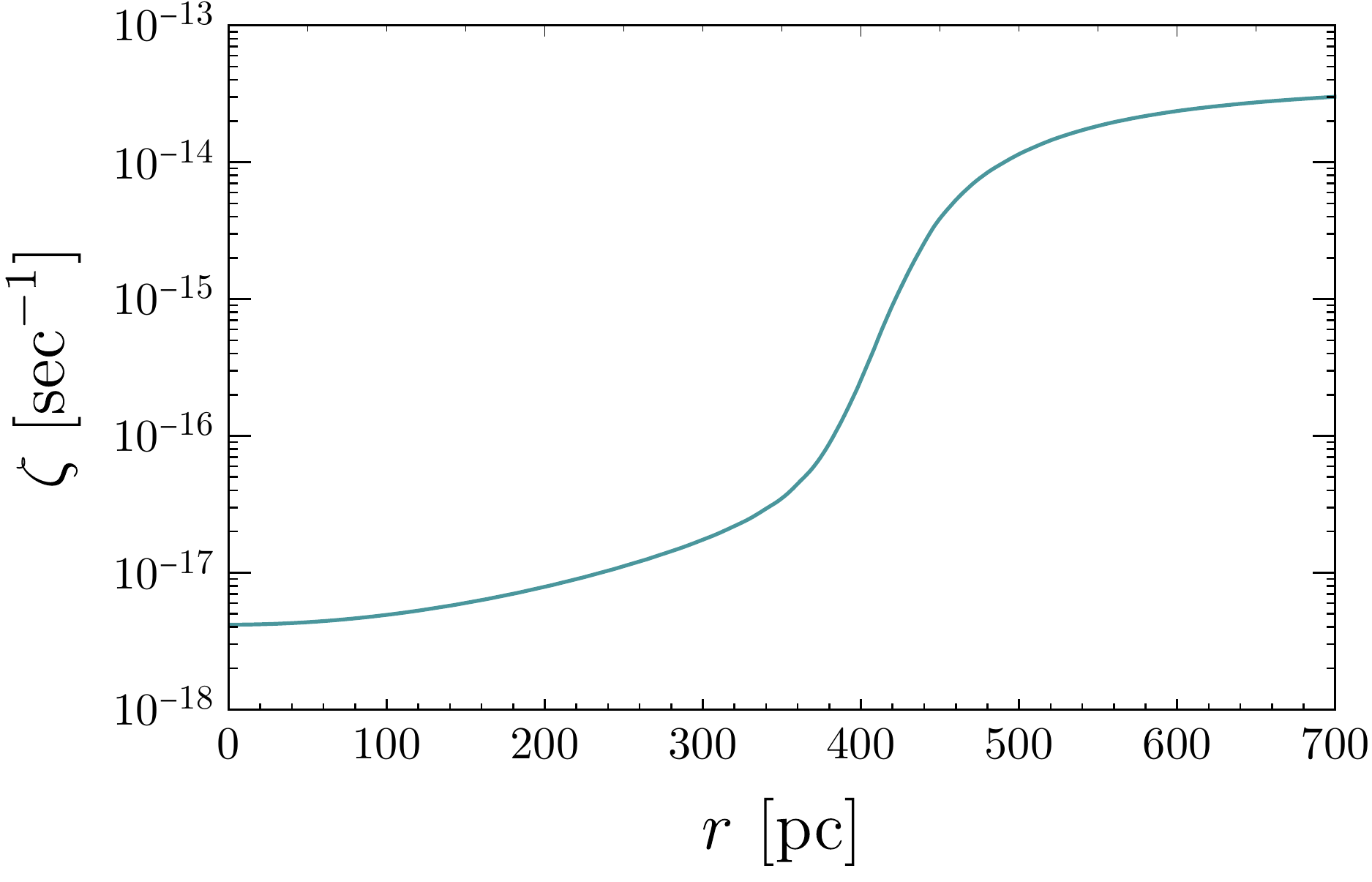}
\caption{ (Top) Profiles of atomic hydrogen and free electrons in Leo T, obtained from the procedure described in the main text. We choose $T_{\rm wnm}= 6000$ K, $M_{\rm wnm} = 2.8\times 10^{5}M_\odot$, $r_{\rm wnm} = 350$ pc, $v_s = 30$ km/sec [scale velocity, defined as $v_s^2 = (4\pi G/3) \rho_s r_s^2$], and $r_s = 709$ pc \citep{RyanWeber:2007fb, Faerman:2013pmm}. This result was also obtained in \citet{Wadekar:2019xnf}, but shown here for the self-contained discussion. (Bottom) The total photoionization rate in Leo T due to cosmic UV background \citep{Sternberg:2002xs, 2012ApJ...746..125H}. }
\label{fig:LeoT_profile}
\end{figure}

The profile of atomic hydrogen and free electrons is needed for computation of the gas cooling rate. 
We assume that the gas is pressure-supported and is confined in the gravitational potential of dark matter halo \citep{Sternberg:2002xs, Faerman:2013pmm}. The ionization structure is obtained by solving ionization equilibrium, $\zeta n(\Hy^0) = \alpha_B n(\Hy^+) n(e^-)$, where $\zeta$ is photoionization rate due to UV background light, and $\alpha_B$ is case B recombination rate (excluding a recombination to the ground state).
For dark matter, Burkert profile is adopted 
\bea
\rho_{\rm dm}(r) = \frac{\rho_s}{(1+r/r_s) ( 1 + (r / r_s)^2)},
\eea
with best fit parameters for observed Leo T HI column density, $r_s = 709$ pc and $\rho_s = 3.8 \GeV/\cm^3$ \citep{Faerman:2013pmm}.\footnote{The cuspy NFW profile might be used.
The best fit parameters for the NFW profile yields the enclosed dark matter mass within 300~pc as $M_{300} \sim 6.5\times 10^6 \, M_\odot$, while the best fit Burkert profile predicts $M_{300} = 8\times 10^6 \, M_\odot$~\citep{Faerman:2013pmm}. 
This might result in $\sim20\%$ change in the PBH heating rate. 
The best fit parameters for the NFW profile, however, is less consistent to the $\Lambda$CDM prediction obtained from large scale N-body simulation Aquarius~\citep{2008MNRAS.391.1685S}.  }
By iteratively solving ionization equilibrium equation while accounting self-shielding of UV background light due to atomic hydrogen, we reproduce the ionization structure of Leo T \citep{Wadekar:2019xnf}. Although the profile of atomic hydrogen and free electrons has already obtained in \citet{Wadekar:2019xnf}, we nevertheless show them in Figure \ref{fig:LeoT_profile} as well as the photoionization rate due to the cosmic UV background for self-contained discussion [cf. Fig.~1 in \citet{Wadekar:2019xnf}].  
We have assumed that cosmic ray ionization has negligible effects on overall ionization structure of Leo T. This will be justified in Section \ref{sec:cosmic_ray}

\section{Cooling and heating}\label{sec:cooling_heating}
The energy transfer rate per unit volume is
\bea
\frac{dE}{dV dt} = \dot{\cal H} - \dot{\cal C},
\eea
where $\dot{\cal H}$ is the total heating rate, and $\dot{\cal C}$ is the total cooling rate. 
We derive a new bound on the fraction of PBH dark matter by imposing
\bea
\frac{1}{V} \int dV \, \dot{\cal H}_{\rm BH} < \frac{1}{V} \int dV\,  \dot{\cal C},
\eea 
where $\dot{\cal H}_{\rm BH}$ is the heating rate due to PBH Hawking radiation. The volume integration is performed over the volume of system under the consideration; in our case, this would the volume of WNM in Leo T. Although there are other heating sources such as cosmic ray and photoionization heating, we ignore them for a conservative estimate. We derive $\dot{\cal C}$ in the next section, $\dot{\cal H}_{\rm BH}$ in Section \ref{sec:heating}, and present the result in Section \ref{sec:result}.

\subsection{Cooling}\label{sec:cooling}
Cooling is a process that extracts energy from the system. For the interstellar medium, it dominated by photon emission. 
A variety of processes contribute to the gas cooling. It includes rotational and vibrational excitations of molecules, fine structure transitions of metals, Lyman-$\alpha$ transition, and bremsstrahlung of ions.
For warm neutral medium of $n_\Hy  \sim 0.1 \cm^{-3}$, $T_{\rm wnm} \simeq 6000$ K and $x_e = n_e / n_\Hy \sim {\cal O}(10^{-2})$, most important coolants are metal line transitions, and perhaps, \lya transition and ro-vibrational excitation of H$_2$ in a metal poor system. 
Among metals, C$^+$, O$^0$, and Si$^+$ lines are dominant coolants at the temperature and electron density of our primary interest. 

Since the cooling takes place through atomic and molecular de-excitation, it can be written as
\bea
\dot{\cal C} = \sum_{u> \ell} n_u A_{u\ell} \omega_{u\ell},
\label{cooling_full}
\eea
where $n_u$ is the density of excited state $u$, $\omega_{u\ell} = \omega_u - \omega_\ell$ is the level spacing, and $A_{u\ell}$ is the spontaneous decay rate. We use the subscript $u$ and $\ell$ to indicate upper and lower atomic levels, respectively. We ignore the stimulated emission since the photon occupation number at any relevant frequencies is smaller than unity. 
This expression assumes that all emitted photons from atomic and molecular de-excitation can escape the system, which is a valid assumption for photons from fine structure transitions in WNM. 
To compute the cooling rate, we will use an approximate expression given in Eq.~\eqref{cooling_approx} instead of Eq.~\eqref{cooling_full}. A derivation is detailed in the next paragraph with two-level system and uninterested readers can directly jump to Eq.~\eqref{cooling_approx}. 

The excited state density $n_u$ is needed to compute $\dot{\cal C}$. Most cooling arises from transitions among a few low-lying states. Consider two-level system for simplicity. The cooling transition in C$^+$, one of the major coolants, can be accurately described by two-level system. 
The population of excited state is obtained from the collisional equilibrium condition,
\bea
\frac{dn_1}{dt} = n_0 C_{01} - n_1 (C_{10} + A_{10}) = 0.
\eea
The collisional de-excitation rate $C_{10}$ is 
\bea
C_{10} = \sum_c n(c) k_{10}(c),
\eea
where $n(c)$ is the number density of collision partner $c$ ($c=e^-,\, \Hy^0, \, p^+, {\rm He}^0, \cdots$), and $k_{10}(c) = \langle \sigma v \rangle_{1\to 0}(c)$ is a thermal-averaged de-excitation rate. The subscript 0 and 1 denote the ground and excited state, respectively.
The collisional excitation and de-excitation rate are related by the detailed balance, $C_{01} = C_{10} (g_1/g_0) e^{-\omega_{10}/T}$. 
The excited population is $n_1 / n_0 = C_{01} / (C_{10} + A_{10})$, and therefore, the cooling rate in the two-level system is
\bea
\dot{\cal C} = n_0 C_{01} \omega_{10} \left[ 1 + \sum_c \frac{n(c)}{n_{\rm crit}(c)} \right]^{-1},
\eea
where we define the critical density as $n_{\rm crit}(c) = A_{10}/k_{10}(c)$; this is the density at which the collisional de-excitation rate and spontaneous decay rate becomes equal.
In the limit $n(c) \ll n_{\rm crit}(c)$, the cooling rate is the same as the collisional excitation rate weighted by the energy level difference. 
For $N$-level system, it is a straightforward to show
\bea
\dot{\cal C} \approx  \sum_u n_0 C_{0 u}\omega_{u0} ,
\label{cooling_approx}
\eea
when $n(c) \ll n_{\rm crit}(c)$. In what follows, we always assume $n(c) \ll n_{\rm crit}(c)$ and use Eq.~\eqref{cooling_approx} to compute the cooling rate.

To grasp an idea of how large the cooling rate is, let us take an example of C$^+$. The strongest cooling line is the 158$\mu$m transition between two lowest-lying states, $^2P_{3/2}\to\,^2P_{1/2}$. 
The energy of photon is $\omega_{10} = 2\pi / (158\mu{\rm m})=  8\times 10^{-3}\eV$, and the spontaneous decay rate is  $A_{10} = 2\times 10^{-6} \,{\rm sec}^{-1}$. 
The rate coefficients at $T = 6000$ K are
\bea
k_{10}(e^-) &=& 5\times 10^{-8} \,{\rm cm^3/sec},
\\
k_{10}(\Hy^0) &=& 2\times 10^{-9} \,{\rm cm^3/sec},
\eea
which are obtained from the fitting formula in \citet{Draine:2317527}. The critical density is $n_{\rm crit}(e^-) = 50 \,{\rm cm^{-3}}$ and $n_{\rm crit}(\Hy^0) = 10^3\,{\rm cm^{-3}}$, confirming $n(c) \ll n_{\rm crit}(c)$. 
The cooling rate is therefore
\bea
\dot{\cal C}_{\rm CII} \simeq n_0 C_{01} \omega_{10}
\sim
6\times 10^{-27} \,{\rm erg/cm^3/sec} 
\times \left(\frac{{\cal A}_C}{10^{-4}}\right) \left( \frac{n_\Hy}{1\,{\rm cm^{-3}}} \right)^2 ,
\label{C_carbon_eq}
\eea
where we have chosen $x_e = x_p = 10^{-2}$, and ${\cal A}_C = n_0 / n_\Hy$ is the carbon abundance normalized with the total hydrogen density. The carbon abundance in the solar neighborhood is ${\cal A}_C \simeq 10^{-4}$.  
The cooling rate from the other metal lines can be computed in the same way, and all of them is proportional to the gas-phase metal abundance ${\cal A}_X$. Due to dust grains in interstellar medium, however, gas-phase metals can condense to a solid form, and therefore, the gas-phase abundance is depleted from the stellar metal abundance. This is called interstellar depletion. A complication related to the gas-phase metallicity will be discussed in Section \ref{sec:discussion}.
We refer readers to \citet{1978ppim.book.....S, 2005pcim.book.....T, Draine:2317527} for general discussions on the cooling and heating of interstellar medium.

We comment on two other cooling sources, which do not directly depend on the metal abundance. This includes ro-vibrational transitions in molecular hydrogen H$_2$, and the Lyman-$\alpha$ transition. 
The cooling rate due to H$_2$ can be parametrized as
\bea
\dot{\cal C}_{\rm H_2} =  \sum_c n(c) n({\rm H_2}) \Lambda_{\rm H_2}(c).
\label{H2cooling}
\eea
Analytic fits for $\Lambda_{\Hy_2}$ are given in \citet{Glover:2008pz,2016arXiv161005679G} for $c = {\rm e},\,{\rm H}^0,\, {\rm H}^+,\, {\rm He}$. The abundance of molecular hydrogen can be obtained by solving chemical network. Instead, we use an analytic expression for H$_2$ abundance obtained from the reduced chemical network \citep{2019ApJ...881..160B}.
On the other hand, the \lya cooling rate is \citep{Dalgarno:1972ak, 1978ppim.book.....S, 1981MNRAS.197..553B}
\bea
\dot{\cal C}_{\rm Ly\alpha} = n_{\rm HI} n_e
\left[ 7.5\times 10^{-19} e^{-118348 \,{\rm K}/T} /(1+T_5^{1/2}) \right] {\rm erg\, cm^{-3}\, s^{-1}},
\label{lya}
\eea
where $T_5 \equiv T / 10^5$ K. The exponent arises from the energy level difference $\omega_{10} \simeq 10.2\eV $. Although the excitation rate is exponentially suppressed, \lya cooling dominates the metal cooling for $T>10^4$ K since the hydrogen is more abundant than metals, and the Lyman-$\alpha$ photon is much more energetic compared to the photon coming from fine structure transitions. The \lya and H$_2$ cooling could be sub-leading or even comparable to the metal cooling in a metal poor environment under a certain condition. In Leo T, we find that \lya and H$_2$ cooling remain as sub-leading contributions to the total cooling rate. We will discuss this in Section \ref{sec:isrf}.

\begin{figure}
\centering
\includegraphics[width=\columnwidth]{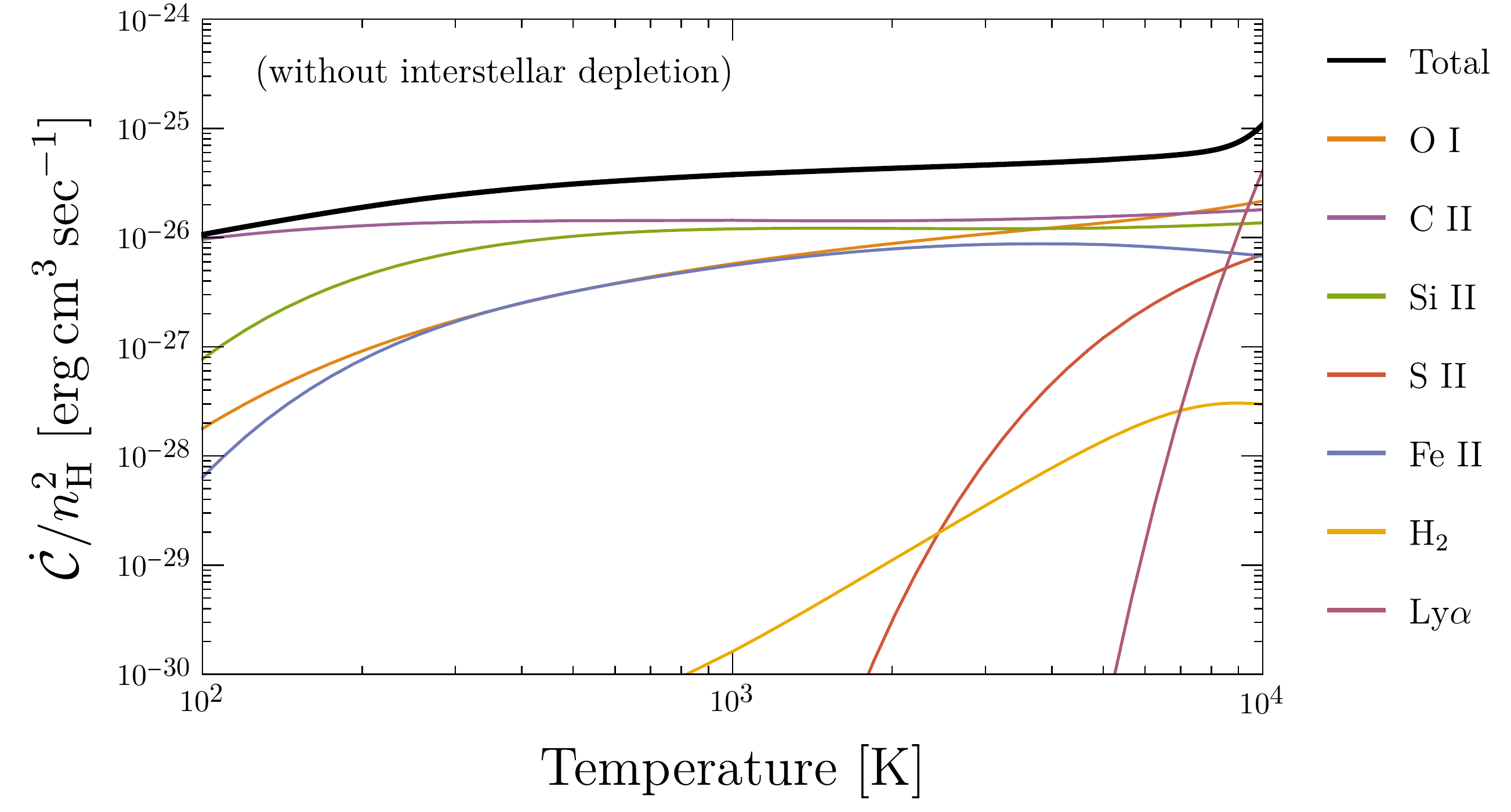}
\includegraphics[width=0.825\columnwidth]{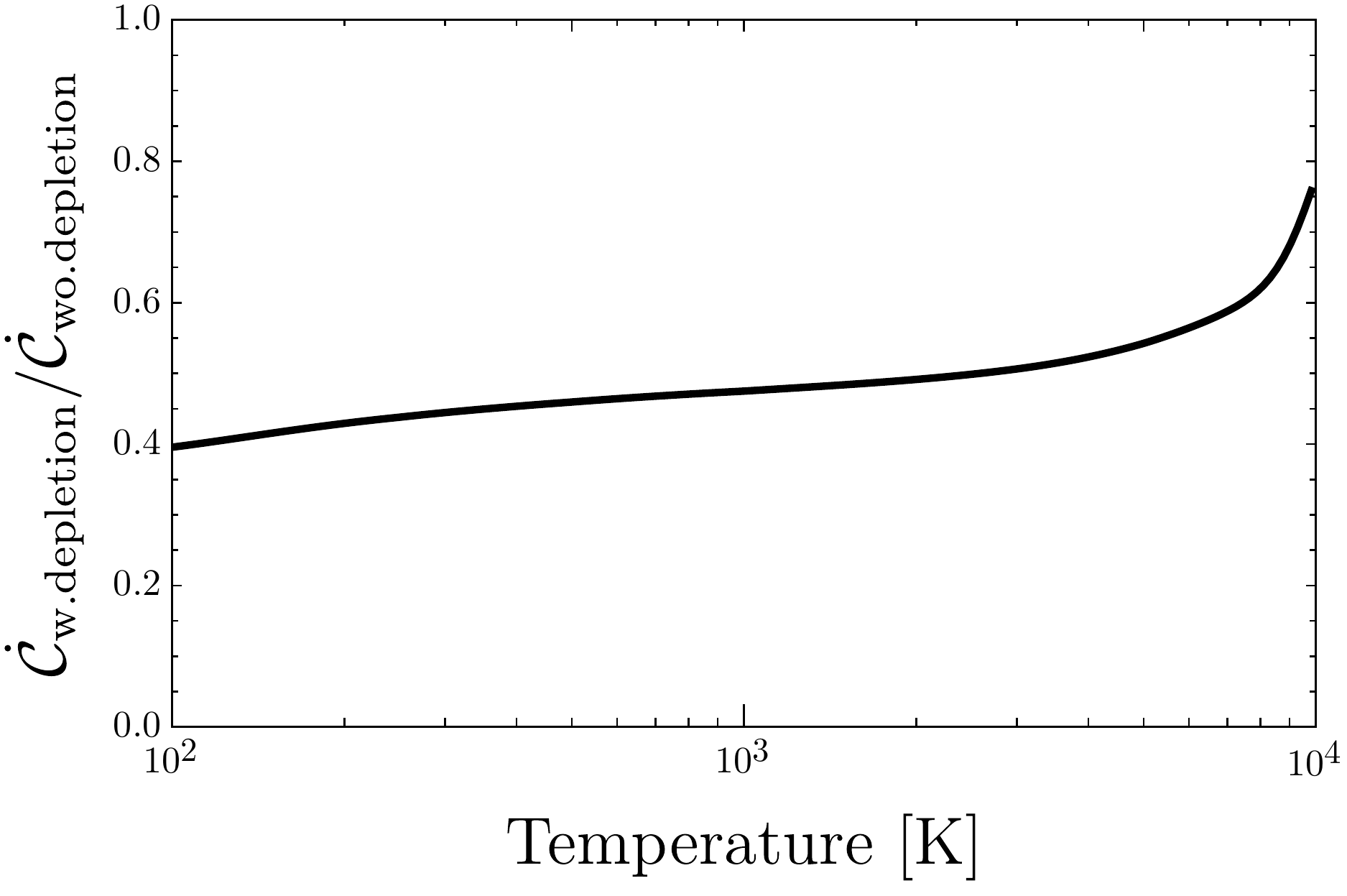}
\qquad\,\,\,\,\,\,
\caption{
(Top) The cooling function at the solar metallicity $Z = Z_\odot$, while the metal abundance is chosen before the interstellar depletion. We choose $x_e = x_p = 10^{-2}$, $x_{\Hy_2} = n(\Hy_2) / n_\Hy= 10^{-6}$, $n_\Hy = 1\cm^{-3}$. 
(Bottom) The ratio between the cooling rate with and without the interstellar depletion. 
Since the depletion of iron in the local interstellar medium is significant ($>$90\%), it could be as important as the other elements in any environment where the depletion can be neglected.
See Table~9.5 in \citet{Draine:2317527} for the effects of depletion of each element on the cooling rate.
We emphasize that the analysis in the main text uses the cooling rate with metal abundance before the interstellar depletion. The cooling rate ratio curve with depleted metal abundance is included only for the completeness of the discussion. 
 See the main text, especially Section~\ref{sec:metallicity}, for details. }
\label{fig:cooling}
\end{figure}

In Figure \ref{fig:cooling}, we show the total cooling rate as well as individual contributions. We show the cooling curves for the solar metal abundance \citep{Asplund:2009fu} without interstellar depletion \citep{2009ApJ...700.1299J, Draine:2317527} as well as the ratio between cooling rate with and without the depletion. For the figure, we choose $x_e = x_p = 10^{-2}$, $x(\Hy_2) = n(\Hy_2)/n_\Hy= 10^{-6}$, and $n_\Hy = 1\cm^{-3}$. 
The electron-impact excitation rate coefficients $k_{u\ell}(e^-)= \langle \sigma v \rangle_{u\to \ell}(e^-)$ are taken from Stout atomic library \citep{2015ApJ...807..118L}.\footnote{The electron-impact excitation rate for each element is originally obtained in the following references: C$^+$ \citep{2008A&A...486..629T}, O$^0$ \citep{1998MNRAS.293L..83B, 2007A&A...462..781B}, Si$^+$ \citep{2014MNRAS.442..388A}, Fe$^+$ \citep{1999ApJS..120..101V}, S$^+$ \citep{2010ApJS..188...32T}.}
Since we are interested in cooling of WNM, hydrogen-impact excitation could be equally important.
We have used \citet{2005ApJ...620..537B} for C$^+$ and Si$^+$ and \citet{2018MNRAS.474.2313L} for O$^0$ for the hydrogen-impact excitation rates $k_{u\ell}(\Hy^0) = \langle \sigma v \rangle_{u\to\ell}(\Hy^0)$. 
Since the quantum mechanical computation of hydrogen-impact excitation of C$^+$ and Si$^+$ is only available for limited temperature range, $15\,{\rm K} \leq T \leq 2000$ K, we extrapolate the rata coefficients to $T_{\rm wnm} = 6000$ K based on the fitting formulae given in \citet{2005ApJ...620..537B}. We use analytic fits given in Appendix \ref{sec:Ofitting} for H$^0$-O$^0$ collision.
Most dramatic change in the cooling rate with and without interstellar depletion arises from Fe II lines since the abundance of iron is depleted more than 90 \% in the local interstellar medium relative to the solar abundance \citep{2009ApJ...700.1299J, Draine:2317527}. The difference in the total cooling rate with and without depletion remains less than a factor three for the shown temperature range.

\subsection{Heating}\label{sec:heating}
We consider the heating from Hawking radiations. 
Primordial black holes could inject energetic particles to the interstellar medium. 
The particle production rate per internal degree of freedom per unit volume is
\bea
d\dot{n} = n_{\rm dm} \frac{(\sigma_{\rm abs}v)}{e^{\omega/T} \pm 1} \frac{d^3k}{(2\pi)^3} = n_{\rm dm} \frac{\Gamma_{\rm abs}}{2\pi} \frac{d\omega}{e^{\omega/T} \pm 1},
\eea
where $\sigma_{\rm abs}$ is an absorption cross section due to the black hole, $\Gamma_{\rm abs} = \sigma_{\rm abs}k^2/\pi$ is the absorption probability of emitted particles, and $n_{\rm dm} = f_{\rm BH}(\rho_{\rm dm} / M)$ is the number density of PBHs with the PBH dark matter fraction $f_{\rm BH}$. We assume that PBHs are Schwarzschild black holes with a $\delta$-function mass distribution.  
The absorption probability can be obtained by numerically solving wave equation under black hole geometries \citep{Page:1976df, Page:1977um, MacGibbon:1990zk}. Instead, we use the publicly available code {\tt blackHawk} \citep{Arbey:2019mbc} to compute the spectrum.

Once electrons are produced, their energy is lost by elastic scatterings with other charged particles and by ionization/excitation of neutral atoms. 
An energy loss due to elastic scatterings increases thermal energy of the medium, and hence, contributes to the heating rate. 
In addition, secondary electrons produced from the ionization of primary electron also contribute to the heating rate. 
Let us define $f_{\rm heat}(\omega)$ as a fraction of primary electron's energy that is deposited as heat.
Then the heating rate due to electrons and positrons is 
\bea
\dot{\cal H}_{\rm BH} = \int d \omega (\omega - m_e) f_{\rm heat}(\omega) \frac{d\dot{n}_{e^\pm}}{d\omega}.
\label{heating_DM}
\eea
The heating rate from Hawking radiation is equal to the convolution of the total kinetic energy carried by emitted electrons and positrons with the heat deposition efficiency. 
This expression assumes that the electrons and positrons are trapped within  WNM at least for a relevant heating time scale. 
This assumption will be justified in Section~\ref{sec:transport}.
For the heat deposition efficiency $f_{\rm heat}(\omega)$, we adopt the analytic fit given in the Appendix of \citet{Ricotti:2001zf} [See their Eqs.~(A3)--(A4) as well as Table~2],
\bea
f_{\rm heat}(\omega) \simeq 1 - (1- x_e^{0.27})^{1.32} + 4x_e^{0.4} (1-x_e^{0.34})^{2}\left( \frac{11\eV}{\omega} \right)^{0.7},
\eea 
for $\omega > 11\eV$, 
which is based on the numerical simulation by \citet{1985ApJ...298..268S}. This fitting formula agrees with numerical results from a more recent Monte Carlo simulation \citep{2010MNRAS.404.1869F}. Above 1 keV, the heat deposition efficiency becomes almost constant, and the energy of primary electron is more or less equally distributed as heat, ionization, and excitation \citep{2010MNRAS.404.1869F}. 

\section{Result}\label{sec:result}
We apply our discussion to the dwarf galaxy Leo T.  For the metal abundance, we take the solar abundance without the interstellar depletion, and rescale them according to the observed metallicity,
\bea
{\cal A}_X = 10^{[\rm X/H]} {\cal A}_{X,\odot} \approx 10^{[\rm Fe/H]} {\cal A}_{X,\odot},
\label{abundance}
\eea
where ${\cal A}_{X,\odot}$ is the solar abundance of metal species $X$ without depletion. We assume $[{\rm X/H}] = [{\rm Fe/H}]$ for all metal elements $X$ under the consideration, i.e. the elemental abundance of the gas in Leo T follows solar abundance pattern. We use $[{\rm Fe/H}] = -1.74$ \citep{2013ApJ...779..102K}. 
Using the profile of atomic hydrogen and electron as an input, we compute the volume-averaged cooling rate as
\bea
\dot{\cal C} \simeq 7 \times 10^{-30} \, {\rm erg/cm^3/sec}.
\label{avg_cooling}
\eea
The averaging is done over the spherical volume of WNM with a radius $r_{\rm wnm} = 350\,{\rm pc}$. 
Note that the averaged cooling rate changes by $\sim 20\%$ when the radius of averaging volume changes by 10\%. 
For the PBH mass smaller than a few times $10^{-17}M_\odot$, we find the average heating rate as
\bea
\dot{\cal H}_{\rm BH} \simeq (7 \times 10^{-27} \,{\rm erg/cm^3/sec}) \times (10^{-17}M_\odot/M)^3.
\eea

By comparing the volume-averaged cooling and heating rate Eq.~\eqref{heating_DM}, we place an upper bound on the PBH dark matter fraction as shown in Figure \ref{fig:result}. Our result based on the heating of ISM from fast electrons are shown as turquoise shaded region. We also show other constraints, arising from extragalactic $\gamma$-ray flux \citep[purple shaded,][]{Ballesteros:2019exr} and Galactic $\gamma$-ray flux \citep[orange shaded,][]{Laha:2020ivk} for comparison. 
The other interesting bounds originating from CMB \citep{Clark:2016nst, Stocker:2018avm, Poulter:2019ooo, Acharya:2019xla, Acharya:2020jbv}, cosmic-ray \citep{Boudaud:2018hqb}, and a future forecast from 21cm absorption \citep{Clark:2018ghm}, are not shown in this figure. 
The constraint is shown for $M \geq 10^{-17} M_\odot$. Below this mass, we expect that the constraint on the PBH dark matter fraction scales as $f_{\rm BH} \propto M^3$ for the following reason. 
For this mass range, Hawking temperature becomes larger than the rest mass of electron, and the heating rate in this case can be approximated as $\dot{\cal H}_{\rm BH} \approx n_{\rm dm} \bar{f}_{\rm heat} P_{e^{\pm}}$ where $\bar{f}_{\rm heat}$ is the asymptotic value of heat deposition fraction at high energy $\sim{\cal O}(0.1)$, and $P_{e^{\pm}}$ is the total power radiated through electrons and positrons from a black hole. The total power can be approximated as $P_{e^\pm} \sim A_{\rm BH} T_{\rm BH}^4 \propto T_{\rm BH}^2$ with a surface area of black hole $A_{\rm BH} = 4\pi r_s^2$, while $n_{\rm dm} = \rho_{\rm dm} / M$. Therefore, $\dot{\cal H}_{\rm BH} \propto T_{\rm BH}^2/M \propto 1/M^3$, and thus, we expect the bound should scale as $f_{\rm BH} \propto M^3$ for smaller mass as long as energy loss time scale is shorter than the cooling time scale and the confinement time scale, which will be discussed in Section \ref{sec:time_scales} and \ref{sec:transport}.

\section{Discussion}\label{sec:discussion}
In this section, we discuss details related to the computation of cooling and heating rate, intended to justify some simplifications and assumptions made. 

\subsection{Metallicity}\label{sec:metallicity}
The cooling is dominated by metal lines. The cooling rate from each metal element is proportional to the gas-phase abundance ${\cal A}_X$. For this, we have used Eq.~\eqref{abundance}, ${\cal A}_X / {\cal A}_{X,\odot} \approx 10^{[{\rm Fe/H}]}$. Two assumptions are made: (i) there is negligible interstellar depletion such that the gas-phase metallicity is similar to the stellar metallicity and (ii) the chemical composition pattern in Leo T follows solar composition pattern, i.e. $[{\rm X/H}]\simeq {\rm constant}$ for all elements under consideration.

Since the depletion of gas-phase elements takes place as a dust grain captures gas-phase element, we naturally expect that the degree of depletion is proportional to the dust abundance. For a system following the solar abundance pattern, the gas-phase abundance can be approximated as \citep{2019ApJ...881..160B}
\bea
{\cal A}_{X} \approx {\cal A}_{X,\odot} (Z/Z_\odot) [ 1 - \delta_X Z_d(Z_\odot/Z) ],
\eea
where $Z_d$ is the dust-to-gas mass ratio normalized as $Z_d = 1$ in the local ISM, and $\delta_X$ is a degree of depletion at the solar metallicity.

The dust-to-gas ratio scales linearly with metallicity, $Z_d \propto Z$, for  systems with near-solar metallicity \citep{1998ApJ...501..643D}. For metal-poor galaxies, recent studies \citep{2014A&A...563A..31R, 2014Natur.505..186F} have shown that the trend of dust-to-gas ratio versus metallicity deviates from this simple linear relation, and in fact, \citet{2014A&A...563A..31R} has found a steeper dependence on metallicity, $Z_d \propto Z^{3.1}$. The transition takes place at $Z /Z_\odot\simeq 0.2 $ \citep{2014A&A...563A..31R}. For the system with $Z/Z_\odot \simeq 0.01 $,  which is commonly observed in ultrafaint dwarf galaxies, we find $Z_d(Z_\odot/Z ) \sim 10^{-3}$. In the local ISM, iron is most significantly depleted, and even in this case, $\delta_{\rm Fe} \sim 0.9$ \citep{2009ApJ...700.1299J}. Therefore, we can safely ignore interstellar depletion in low-metallicity systems like Leo T.

Another assumption made in Eq.~\eqref{abundance} is that the elemental composition in Leo T is the same as solar composition pattern. In the other words, we have assumed that $[{\rm X/H}] = \log_{10}(n_X/n_\Hy)_{\rm Leo T} - \log_{10}(n_X/n_\Hy)_{\odot} = {\rm constant}$ for all $X$, indicating that the metal abundance of any given metal species in Leo T is suppressed by the common factor relative to the solar abundance of $X$.  
This assumption is not necessarily true for certain systems. The observations of ultrafaint dwarf galaxies have revealed that the alpha-elements, such as O and Si, are enhanced relative to the solar abundance pattern, i.e. $[\alpha/{\rm Fe}] > 0$. 

This enhancement in $\alpha$-elements is the result of the chemical enrichment due to Type II supernovae (SNe) explosions. It is well known that Type II SNe explosions produce a large amount of $\alpha$-elements, resulting in $[\alpha/{\rm Fe}] > 0$, while Type Ia SNe explosions mainly produce iron-peak elements, and thus, raising $[{\rm Fe/H}]$ and lowering $[\alpha/{\rm Fe}]$. At the same time, a minimum time of $\sim 100$ Myr is required \citep{2008PASJ...60.1327T, 2012MNRAS.426.3282M} for Type Ia SNe to explode, and therefore, we may understand the enhancement in $\alpha$-elements as an indication of star formation lasted less than 100 Myr. From the study of seven ultrafaint dwarf galaxies, \citet{2013ApJ...767..134V} found a sharp drop of $[\alpha/{\rm Fe}]$ around $[{\rm Fe/H}] \sim -2.3$. See also \citet{2015ApJ...799L..21W} for a model of chemical evolution of ultrafaint dwarf galaxies and \citet{2017ApJ...848...85J} for numerical simulation. Since Leo T has a relatively large metallicity $[{\rm Fe/H}] = -1.74$ \citep{2013ApJ...779..102K} and also hosts a recent star formation \citep{2007ApJ...656L..13I, 2008ApJ...680.1112D, 2012ApJ...748...88W}, it is reasonable to expect the elemental abundance of Leo T closely follows solar abundance pattern, $[\alpha/{\rm Fe} ] \simeq 0$. See the recent review by \citet{2019ARA&A..57..375S} for further discussions.

Finally, we comment on a numerical difference in the cooling rate compared to the previous works \citep{Wadekar:2019xnf, Lu:2020bmd}. We have found a factor three larger cooling rate compared to what was found in \citet{Wadekar:2019xnf}. This numerical difference can be attributed to the metallicity used in the computation of the cooling rate as well as the effect of interstellar depletion. For the metallicity used in \citet{Wadekar:2019xnf} ($[{\rm Fe/H}] = -1.99$, \citet{Kirby:2010wd}) and for the depleted elemental abundance, we obtain the averaged cooling rate reasonably close to the one presented in \citet{Wadekar:2019xnf} ($\dot{\cal C} \sim 2\times 10^{-30} \,{\rm erg/cm^3/sec}$).

\subsection{Cosmic ray ionization}\label{sec:cosmic_ray}
When determining the ionization structure of Leo T, we have only accounted ionization due to cosmic UV background, while ignoring a potential contribution from cosmic rays. 
In order to justify this, we have to make sure that the cosmic ray ionization rate is smaller than the average ionization rate due to cosmic UV background $\langle \zeta \rangle \sim 2\times 10^{-17}\,{\rm sec}^{-1}$, obtained from the bottom panel of Figure~\ref{fig:LeoT_profile}. 

The cosmic ray ionization produces secondary electrons with energies $\sim 30\eV$ \citep{Dalgarno:1972ak}. Similar to fast electrons from Hawking radiation, this secondary electron could also contribute to the heating of ISM.
The heating rate is \citep{Dalgarno:1972ak}
\bea
\dot{\cal H}_{\rm cr} 
\!\!\!\!\!&=&\!\!\!\!\! \big[ n(\Hy^0) + n({\rm He}^0) \big] \zeta_{\rm cr} E_h
\nonumber\\
\!\!\!\!\!&\simeq&\!\!\!\!\!
10^{-27} \,{\rm erg/s} 
\times \zeta_{-16}
\big[ n(\Hy^0) + n({\rm He}^0) \big] 
\Big[ 1 + 4 \Big( \frac{x_e}{x_e + 0.07} \Big)^{\frac{1}{2}} \Big]
\nonumber \\
\!\!\!\!\!&\to&\!\!\!\!\! (2 \times 10^{-28} \,{\rm erg/cm^3/sec}) \,\times \zeta_{-16},
\eea
where $E_h$ is the average heat per ionization, and $\zeta_{-16} \equiv \zeta_{\rm cr} / 10^{-16}\sec^{-1}$. The third line is obtained after averaging over the volume of WNM. By comparing averaged cooling rate Eq.~\eqref{avg_cooling}, we find $\zeta_{\rm cr} < 3\times 10^{-18} \sec^{-1}$. 
Since the electron density is determined by the ionization equilibrium condition, $\zeta n({\rm H}^0) = n({\rm H}^+) n(e^-) \alpha_B$, ($\alpha_B$ is case B recombination rate), the ionization fraction is proportional to the square root of total photoionization rate, $x_e \simeq (\zeta / n_H \alpha_B)^{1/2}$. 
An uncertainty due to the cosmic-ray ionization on the cooling rate is therefore estimated to be less than $10\%$.

\subsection{Interstellar radiation field}\label{sec:isrf}
As we have briefly mentioned in Section \ref{sec:cooling}, the cooling due to molecular hydrogen may be comparable to the metal cooling in low metallicity environment because it does not directly depend on the metallicity. This can be  seen from the cooling curves in Figure \ref{fig:cooling}. On the other hand, the abundance of H$_2$ is determined by a network of chemical reactions.
Two important chemical reactions that suppress the abundance of H$_2$ are the photodissociation of molecular hydrogen, $\Hy_2 + \gamma \to \Hy + \Hy$, and the photodetachment of hydrogen ion, $\Hy^- + \gamma \to \Hy + e$, all due to UV interstellar radiation field (ISRF). Therefore, the strength of ISRF is crucial to determine $n(\Hy_2)$, and thus, the cooling due to molecular hydrogen.

To infer the strength of ISRF, we make use of equipartition of energy. In the local ISM, the energy density stored in thermal gas and in starlight are surprisingly similar; $\rho_{\rm thermal} \sim \rho_{\rm starlight} (\omega < 13.6\eV) \sim 0.5 \eV/\cm^3$ \citep{Jenkins:2011ds, Draine:2317527}. On the other hand, the average thermal energy of WNM in Leo T is $\rho_{\rm thermal} = \frac{3}{2} \langle n_{\rm HI} \rangle T_{\rm wnm} \simeq 0.05 \eV/\cm^3$, ten times smaller than the kinetic energy density in the local ISM. We assume that the strength of ISRF is suppressed by the same amount, while the spectral shape remains the same. 
In such case, the cooling due to molecular hydrogen remains sub-dominant. For stronger ISRF, the H$_2$ cooling becomes negligible since H$_2$ abundance is more strongly suppressed, while for an order of magnitude weaker ISRF strength relative to the value inferred from equipartition, the total averaged cooling rate increases by $\sim$ 50\%, signaling that the H$_2$ cooling becomes comparable to the metal cooling rate. 
Therefore, our result remains intact for a wide range of ISRF strength unless it is smaller by more an order of magnitude than what is inferred from equipartition theorem.
We also note that the spectrum of ISRF in metal poor galaxies is harder compared to Galactic ISRF \citep{2006A&A...446..877M}, which may further suppress H$_2$ abundance.

\subsection{Photoelectric heating}
Can photons from Hawking evaporation provide additional heating? Photons could also heat ISM by producing photoelectrons. The heating rate can be computed as
\bea
\dot{\cal H} \sim n_{\rm HI} \int d\omega \, (\omega - I_\Hy) \sigma_\Hy(\omega) f_{\rm heat}(\omega)\frac{dn_\gamma}{d\omega},
\eea
where $I_\Hy=13.6\eV$ is the ionization potential of hydrogen, $\sigma_\Hy$ is photoionization cross section, and $dn_\gamma/d\omega$ is the photon number density in $[\omega,\omega+d\omega]$ frequency interval. 
The photoelectric cross section above keV scales as $\sigma_\Hy \propto \omega^{-3.5}$, while $dn_\gamma/d\omega$ at $\omega < T_{\rm BH}$ scales as $dn_\gamma/d\omega \propto \omega^4$ due to the gray body factor for spin one bosons \citep{Page:1976df}. This allows us to estimate the heating rate by evaluating the integrand at the peak of $dn_\gamma/d\omega$, which is $\omega \sim 2\pi T_{\rm BH}$. Using $\sigma_\Hy \simeq 5\times 10^{-17}\cm^2 (I_\Hy/\omega)^{3.5}$, and $dn_\gamma/d\omega = N_{\rm dm} (dN_\gamma/d\omega dt)$ with dark matter column density $N_{\rm dm} \sim n_{\rm dm} r_s$ and $(dN_\gamma/d\omega dt)_{\rm peak} \simeq 10^{21} \GeV^{-1}\sec^{-1}$ \citep{MacGibbon:1990zk}, we find $\dot{\cal H} \sim (10^{-40}\,{\rm erg/cm^3/sec})\times (n_{\rm HI}/1\cm^{-3}) (M/ 10^{-16}M_\odot)^{1/2}$. Therefore, photons from Hawking evaporation have negligible heating effects.

\subsection{Time scales}\label{sec:time_scales}
Our result is based on the heating rate Eq.~\eqref{heating_DM}. This expression assumes that the energy of primary electron is instantaneously injected into the ISM as a thermal energy. This assumption is justified only if the thermalization time scale is shorter than the cooling time scale. The cooling time scale is 
\bea
t_{\rm cool} = \frac{E_{\rm th} n}{\dot{\cal C}} \simeq 0.4 \, {\rm Gyr},
\eea
where we have used the average gas density $n = 0.07 \cm^{-3}$ in WNM, the average cooling rate Eq.~\eqref{avg_cooling}, and the thermal energy $E_{\rm th} = (3/2) T_{\rm wnm}$. On the other hand, the thermalization time scale can be obtained from the following electron energy evolution equation,
\bea
\frac{dE}{dt} \simeq 
- \left( \frac{\partial E}{\partial t} \right)_{\rm el} 
- \left( \frac{\partial E}{\partial t} \right)_{\rm ion} 
+ \cdots,
\eea
where $(\partial E/\partial t)_{\rm el}$ and $(\partial E / \partial t)_{\rm ion}$ are the energy loss due to elastic scattering and ionization, respectively. Thermalization time scale through the elastic scattering can be defined as
\bea
t_{\rm th} &=& \frac{E}{(\partial E / \partial t)_{\rm el}} \simeq \frac{2}{n_e \sigma_{\rm tr}v_{\rm rel}}
\nonumber\\
&\sim& (0.03 \,{\rm Gyr}) \times \left(\frac{E}{{\rm MeV}}\right) \left(\frac{1.5\times 10^{-3}\cm^{-3}}{n_e} \right)
\label{th_time},
\eea
where $\sigma_{\rm tr} \equiv \int d\Omega ( 1 - \cos\theta) d\sigma / d\Omega \simeq 8\pi \alpha^2 \ln \Lambda/ (m_e E) $ is the transport (momentum transfer) cross section in the relativistic regime, $E$ is the kinetic energy, and $\ln \Lambda \simeq \ln (2 p_{\rm CM} v / \omega_p)$ is the Coulomb logarithm with the plasma frequency $\omega_p = (4\pi\alpha n_e/m_e)^{1/2}$ \citep{1971P&SS...19..113S}. The above expression is valid for the relativistic regime. In the nonrelativistic regime, the thermalization time scale scales as $t_{\rm th} \propto E^{3/2}$. Note that the average kinetic energy of electrons from PBH is $\langle E \rangle \sim 3 T_{\rm BH} \simeq 2 \MeV (10^{-17}M_\odot/ M)$. Therefore, elastic scattering is sufficiently strong such that it thermalizes MeV scale energy within the cooling time scale. 

Not only the elastic scattering, but also ionization contributes to thermalization in a significant way \citep{1985ApJ...298..268S}. Each ionization process divides the energy of primary electron into smaller secondary electron energies. The mean energy of secondary electron is $\sim{\cal O}(10)$ eV for MeV scale primary electron \citep{1979ApJ...234..761S}, and thus, the secondary electron can be instantaneously thermalized on time scale of interest. 
The time scale of ionization energy loss is
\bea
t_{\rm ion} = \frac{E}{(\partial E / \partial t)_{\rm ion}} \sim  (3\times 10^{-3} \,{\rm Gyr}) \times \left( \frac{0.06\cm^{-3}}{n_{\rm HI}} \right) \left(\frac{E}{{\rm MeV}}\right) , 
\label{ion_time_scale}
\eea
where the ionization energy loss rate is~\citep{Berestetsky:1982aq}
\bea
\left( \frac{\partial E}{\partial t} \right)_{\rm ion} \!\!\! &=& \!\!\!
\frac{2\pi \alpha_{\rm em}^2 n_{\rm HI}}{m_e v} 
\bigg[
\ln  \frac{m_e^2 (\gamma^2-1)(\gamma-1)}{2I_H^2} 
- \left( \frac{2}{\gamma} - \frac{1}{\gamma^2} \right) \ln 2
\nonumber\\
&&
+ \frac{1}{\gamma^2}
+ \frac{(\gamma-1)^2}{8\gamma^2} 
\bigg]
\label{ion_loss}
\\
\!\!\! &\simeq&\!\!\!
\frac{2\pi \alpha^2 n_{\rm HI} }{ m_e} \ln \frac{m_e^2 \gamma^3}{2I_\Hy^2}. 
\eea
Here, the second line is obtained in the relativistic regime.
As long as $M\gtrsim 10^{-19}M_\odot$, the above time scale is much shorter than the cooling time scale, and thus, we can assume that a significant fraction of energy is thermalized before the cooling is initiated.

\subsection{Transport of charged particles}\label{sec:transport}
We have computed the PBH heating rate based on Eq.~\eqref{heating_DM}.
This assumes that most of electrons are trapped within WNM in Leo T, at least for the heating time scale of ${\cal O}(1\textrm{--}10)$~Myr. 

To justify the assumption, we first recall the transport of cosmic-ray in our galaxy as it shares similar features with the problem of the electron transport in Leo T. 
The cosmic-ray propagates almost at the speed of light along the magnetic field lines. If the motion were ballistic, the cosmic rays escape our galaxy within a time scale of $\tau_e \sim L / c = 3\times 10^3\,{\rm yr} \times (L/{\rm kpc})$. 
Contrary to this naive estimation, an actual confinement time scale inferred from cosmic-ray clocks indicates $\tau_e \sim 10^7\, {\rm yr}$~ \citep{2001ApJ...563..768Y}, which is almost four orders of magnitude larger than the naive time scale. 
Together with the observed highly-isotropic cosmic-ray distribution, it indicates that cosmic-ray should be diffusively transported.
The origin of diffusive transport is the interaction between cosmic rays and magnetic field fluctuations, especially those generated by cosmic-ray themselves~\citep{1969ApJ...156..445K}. 
This wave-particle interaction suppresses the cosmic-ray streaming velocity to the level of \alf velocity, $v_A \sim 10^{-4}$, so that cosmic-ray effectively travels at the \alf speed not at the speed of light, and therefore, the confinement time scale is dramatically suppressed, $\tau_e \sim L / v_A = 10^7\,{\rm yr} \times (L/{\rm kpc}) (10^{-4} / v_A)$ [see reviews by~\citet{1974ARA&A..12...71W, 1980ARA&A..18..289C} for details]. 

In Leo T, the same is expected to happen for electrons and positrons. 
If the electron mean free path were larger than the size of the system, the distribution of relativistic electrons would be highly anisotropic. 
This leads to a rapid production of hydromagnetic waves, such as \alf and magnetoacoustic waves, through the streaming instability.
Such waves then scatter with electrons and positrons, isotropizing the electron/positron distribution, and reducing the streaming velocity, or equivalently, decreasing the mean free path. 
The streaming velocity would decrease to the level where the production rate becomes equal to damping rates.

\begin{figure}[t]
\centering
~\includegraphics[width=0.94\columnwidth]{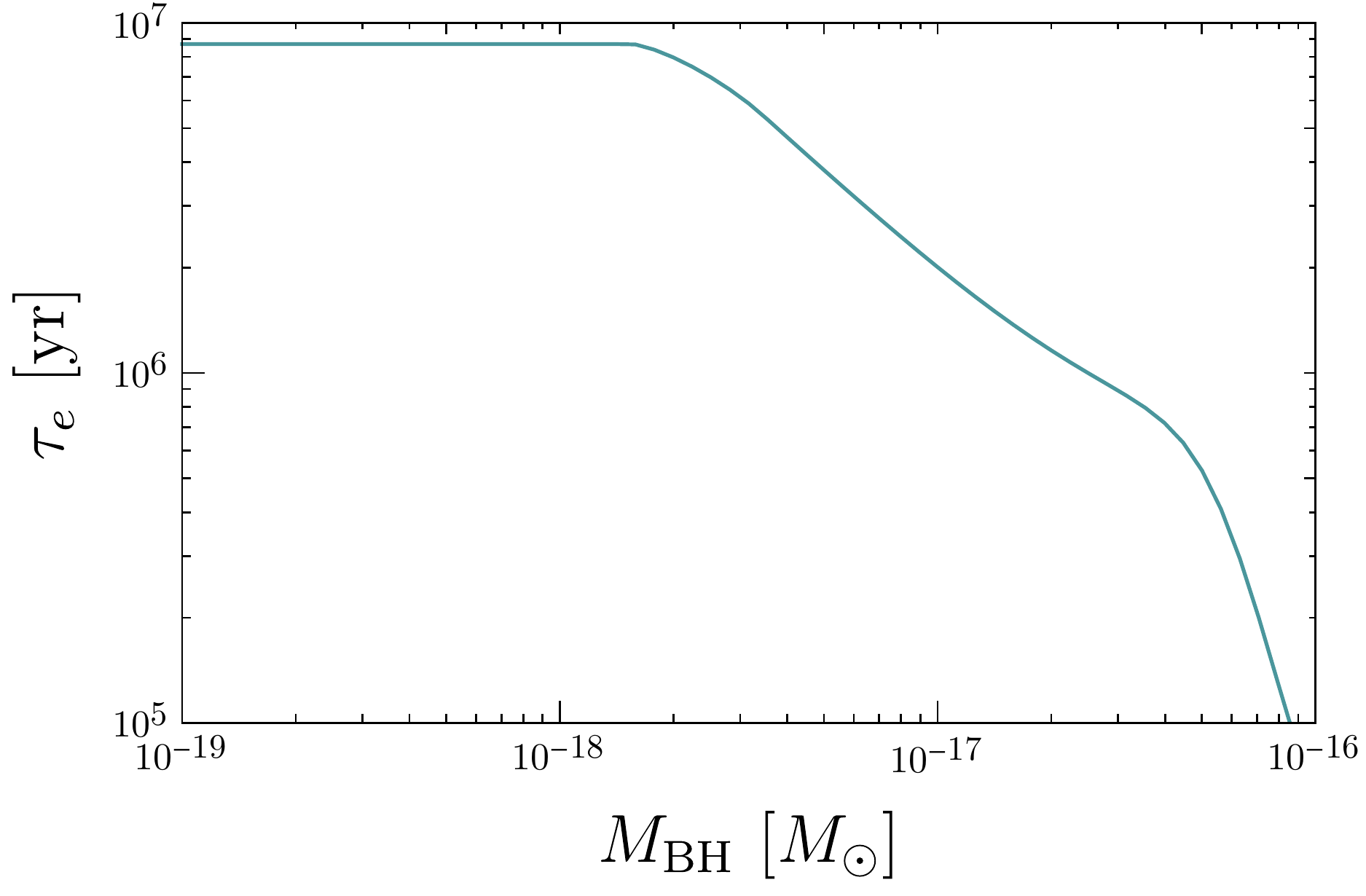}
\includegraphics[width=0.95\columnwidth]{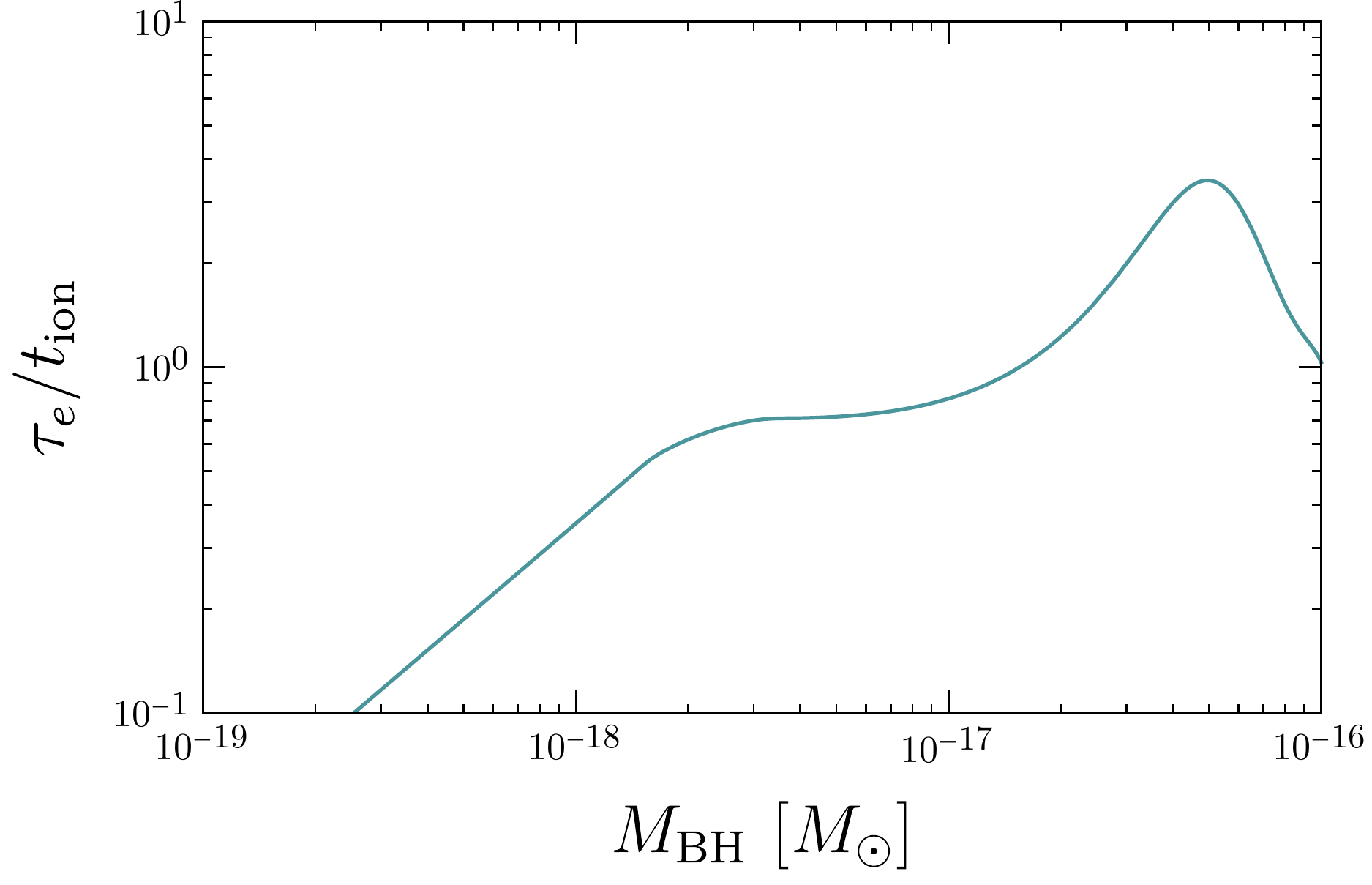}
\caption{(Top) the confinement time scale, defined in Eq.~\eqref{conf_time}. 
Due to the wave-particle interaction, the confinement time scale is ${\cal O}(10^{3\textrm{--}4})$ larger than the naive estimation $\tau_{e}^{\rm naive} = r_{\rm wnm} / c \sim 10^3\,{\rm yr}$. 
(Bottom) the ratio between the confinement time scale and the ionization energy loss time scale. 
For the ionization energy loss time scale, we numerically solve Eq.~\eqref{ion_loss} to obtain the time scale at which initial particle of kinetic energy is decreased by half. 
The confinement time scale and energy loss time scale is comparable as long as $M\gtrsim 10^{-18}M_\odot$.
For this figure, we have assumed $B_0 = 10^{-6}\,{\rm G}$. }
\label{fig:confinement_time}
\end{figure}

To estimate the confinement time scale in WNM, we compare the wave production rate and a damping rate. In predominantly neutral medium, the dominant source of damping arises from ion-neutral collisions~\citep{1969ApJ...156..445K}.
The ion-neutral damping rate is~\citep{1988ApJ...332..984F}
\bea
\Gamma_{\rm d} \simeq  10^{-9} \, {\rm s}^{-1} \times
( 1 -0.9 x_p )
\left( \frac{n_{\rm HI}}{1\,{\rm cm}^{-3}} \right),
\eea
for $x_p \gtrsim 10^{-4}$. 
On the other hand, the growth rate of \alf wave due to relativistically streaming electron or positron is
\bea
\Gamma_{\rm p}^A \sim 10^{-2} \frac{v_A v_{\rm str}^2}{|k_{||}|} (n_{\rm DM} r_s).
\eea
where $k_{||}$ is the wave vector along the direction of external magnetic field, the \alf velocity is
\bea
v_A = \frac{B}{\sqrt{\rho_{\rm ion}}}
\eea
and the streaming velocity is defined as $\vec{v}_{\rm str} = \int\frac{d^3p}{(2\pi)^3} \vec{v} f_e  / \int \frac{d^3p}{(2\pi)^3} f_e$.
Here, $B$ is strength of magnetic field, and $\rho_{\rm ion}$ is ion mass density.
See Appendix~\ref{app:growth} for a complete expression for the growth rate for hydromagnetic waves.
The critical streaming velocity is obtained by solving $\Gamma_d = \Gamma_p^A$. 
For the relativistic electrons, we find the critical streaming velocity as
\bea
v_{\rm str, c} \sim 5 \left[ 
\frac{\Gamma_d}{p} 
\frac{\sqrt{m_p n(H^+)}}{n_{\rm dm}r_s}
\right]^{1/2},
\label{crit_streaming}
\eea
where $m_p$ is the proton mass.
For this expression, we have used the resonance condition for electron-\alf wave interaction, $k_{||} v_{||} = \Omega_e / \gamma$ with the electron gyro-frequency $\Omega_e = e B / m_e$ and the Lorentz boost factor $\gamma$, and ignored trigonometric factors. 
If the critical streaming velocity becomes smaller than the \alf velocity, the net streaming velocity saturates to the \alf velocity. 
Therefore, the streaming velocity can be written as
\bea
v_{\rm str} = \max\Big[ v_{\rm str,c}, \frac{2}{3} v_A \Big].
\eea
We also define the confinement time scale as 
\bea
\tau_e = \int_0^{r_{\rm wnm}} dr /v_{\rm str}.
\label{conf_time}
\eea

The critical streaming velocity Eq.~\eqref{crit_streaming} scales as $v_{\rm str,c} \propto M$ for the average particle momentum $p = \langle p\rangle$. 
At small black hole masses, it becomes smaller than the \alf velocity, and the streaming velocity saturates to the \alf velocity.
On the other hand, the value of \alf velocity is a function of the background magnetic field. 
While there is known no direct measurement of magnetic field in Leo T, we may estimate the background magnetic field by equipartition of energy,  $B_0 = \sqrt{2 \rho_{\rm thermal}} \simeq 10^{-6}\,{\rm G}$. 
We may also use the observed correlation between star formation rate and the strength of magnetic field in dwarf irregulars~\citep{2011A&A...529A..94C}. 
Combined with star formation rate in Leo T, SFR$\sim 10^{-5} M_\odot/{\rm yr}$~\citep{2012ApJ...748...88W}, the strength of magnetic field is inferred as $B_0 \simeq 0.6\times 10^{-6} \,{\rm G}$, which is similar to the value inferred from equipartition theorem. 
Therefore, we use $B_0 = 10^{-6}\,{\rm G}$ for the analysis.
We note that smaller magnetic field strength leads to a smaller \alf velocity, and it increases the confinement time scale of electrons and positrons produced from PBH of mass $M \lesssim 10^{-18}M_\odot$. 

We show the confinement time scale in Figure~\ref{fig:confinement_time}.
We see that it is three to four orders of magnitude larger than the would-be escape time scale if particles were free-streaming. 
The confinement time scale depends on the momentum of particle, and for the figures, we have chosen the average momentum of particle emitted from the black hole, $p = \langle p \rangle \sim 3\textrm{--}4 \, T_{\rm BH}$.
For the \alf velocity, we take $B_0 = 10^{-6}\,{\rm G}$. 
The streaming velocity saturates to the \alf velocity below $10^{-18}M_\odot$, and the confinement time scale becomes constant. 
On the other hand, the ionization loss time scale increases linearly with the average particle momentum.
We see that the confinement time scale becomes an order of magnitude smaller than the ionization energy loss time scale around $M_{\rm BH} \sim 3 \times 10^{-19} M_\odot$. 
Since the confinement time scale is comparable to the ionization energy loss time scale for $M_{\rm BH}\gtrsim 10^{-18}M_\odot$, we expect most of electrons and positrons deposit their energy into ISM while they are trapped in WNM; our result in Figure~\ref{fig:result}  and the scaling behavior of the bound $f_{\rm BH} \propto M^3$ are valid as long as $M_{\rm BH} \gtrsim 10^{-18}M_\odot$.\footnote{We note that a recent study by \citet{Laha:2020vhg} found a weaker constraint on $f_{\rm BH}$ from Leo T. Authors assumed that warm neutral medium in Leo T is optically thin to relativistic electrons, and they obtained a weaker constraint as most of electrons and positrons escape the system before they deposit a substantial fraction of their energies to the gas.}

\section{Summary}
We have considered the interstellar medium as a probe of primordial black hole dark matter. We have computed the total cooling rate in Leo T, considering various processes such as metal line transitions, transitions in molecular hydrogen levels, and the \lya transition. By comparing the cooling rate with the heating rate due to electrons emitted from PBHs, we have obtained an upper limit on the PBH dark matter fraction $f_{\rm BH}$, that is competitive with or stronger than other existing constraints for $M\lesssim 5\times 10^{-17}M_\odot$ (See Figure \ref{fig:result}). 

We have demonstrated with Leo T that the interstellar medium temperature measurement can be used to constrain low mass PBHs. 
While we have only focused on WNM in the dwarf galaxy Leo T, the same analysis might apply to the other astrophysical system. 
As discussed in previous sections, what makes Leo T particularly susceptible to the additional heating from PBHs is that it hosts a large amount of warm neutral hydrogen gas, and that the metal abundance is low, almost two orders of magnitude low compared to local ISM, so that the cooling rate is suppressed accordingly. 
An interesting direction for a future study would be to find out other astrophysical systems like Leo T and to use them to constrain the parameter space of PBHs and possibly other DM models.

\section*{Acknowledgements}
We would like to thank Kfir Blum, Joshua Eby, Fatemeh Elahi, and Sarah Schon for useful discussions. 
The author especially thanks Kfir Blum and Joshua Eby for helpful suggestions and comments on the first version of the manuscript, Steven R. Furlanetto for providing his numerical results on the energy deposition fraction of fast electrons, and Fran{\c c}ois Lique for the clarification on the hydrogen-impact collision data of atomic oxygen. 
We also thank Ranjan Laha for a private correspondence.
We especially thank anonymous referee for critical comments on the mean free path of electrons as well as various suggestions on the manuscript.
The author thanks the GGI Institute for Theoretical Physics and CERN for hospitality during ``Next Frontiers in the Search for Dark Matter'' and ``Axions in the Lab and in the Cosmos'' workshops, during which this work was conceived. 
This research is supported by the Deutsche Forschungsgemeinschaft under Germany’s Excellence Strategy - EXC 2121 Quantum Universe -
390833306.

\section*{Data Availability}
The numerical results presented in this article are available upon request.

\appendix
\section{Analytic fits for the hydrogen-impact excitation of atomic oxygen}\label{sec:Ofitting}

For the hydrogen-impact collision of atomic oxygen, we have used LAMDA database \citep{2005A&A...432..369S}.\footnote{See \url{https://home.strw.leidenuniv.nl/~moldata/}.} 
The collision data has been obtained in \citet{2018MNRAS.474.2313L}.
The data is obtained for $10$ K -- $8000$ K and can be fitted with
\bea 
k_{21}(H^0) \!\!\!\! &=& \!\!\!\! 1.087 \times 10^{-10}   \Big[ 1 + 2.016 \times T_3^{1.097 -0.129 \ln T_3} \Big]\, {\rm cm^3/sec},
\nonumber\\
k_{20}(H^0) \!\!\!\! &=& \!\!\!\!  4.933 \times 10^{-11} \Big[1 + 7.368 \times T_3^{0.599 -0.064 \ln T_3} \Big] \,{\rm cm^3/sec},
\nonumber\\
k_{10}(H^0) \!\!\!\! &=& \!\!\!\!1.068 \times 10^{-8} \Big[ 1 -0.950 \times T_3^{-0.023 - 0.003 \ln T_3} \Big] \, {\rm cm^3/sec}.
\nonumber
\eea
For $1000\,{\rm K} < T < 8000$ K, the error remains less than three percent. We use the above fitting formula to compute the cooling rate due to hydrogen-impact excitation of atomic oxygen.

\section{Electron \& Positron phase distribution function}\label{app:distribution}
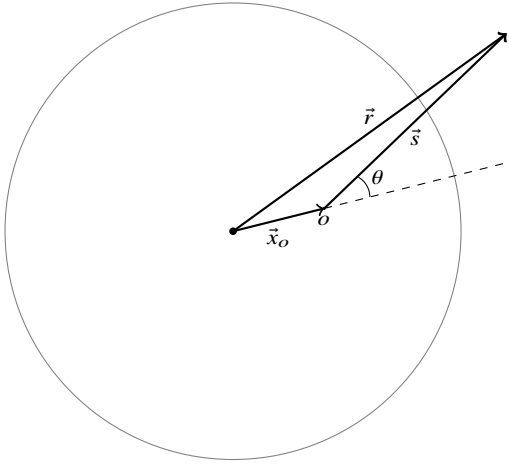
\begin{figure}
\centering
\begin{tikzpicture}
\coordinate (C) at (0,0);
\coordinate (O) at (1.2, 0.3);
\coordinate (S) at (3.6,2.6);
\coordinate (Oe) at (3.6,0.9);

\draw[gray] (C) circle [radius=3cm];

\draw[->, thick] (C) -- (O) node [below, midway] {$\vec{x}_o$};
\draw[->, thick] (C) -- (S) node [above, midway] {$\vec{r}$};
\draw[->, thick] (O) -- (S) node [below,midway] {$\vec{s}$};
\draw[dashed] (O) -- (Oe);

\draw (1.8,0.45) arc [start angle =0, end angle = 60, radius =3mm] node [right] {$\,\theta$};

\fill [black] (C) circle [radius=0.05cm];
\fill [black] (O) circle [radius=0.02cm]  node [black,below] {$o$};

\end{tikzpicture}
\caption{
A coordinate system used to describe the phase space distribution Eq.~\eqref{fe}. 
The electron density at the position $o$ is obtained by integrating the transport equation Eq.~\eqref{stationary}. }
\label{fig:schematic}
\end{figure}

Consider the kinetic equation of the electron,
\bea
\frac{df_e}{dt} = j_{\rm source} - \alpha f_e,
\eea
where the source term due to PBHs is 
$$j_{\rm source } = g_e n_{\rm dm} \frac{\sigma_{\rm abs}v}{e^{\omega/T} +1}$$
with internal degrees of freedom $g_e =2$, and $\alpha$ describes the scattering of electrons with particles in the medium and magnetic field fluctuations. 
For the stationary distribution, the kinetic equation becomes
\bea
\vec{v} \cdot \frac{\partial f_e}{\partial \vec{x}} = j_{\rm source} - \alpha f_e,
\label{stationary}
\eea
and the formal solution is obtained as
\bea
f_e(\vec{x}_o, p) &=& \frac{g_e \sigma_{\rm abs}}{e^{\omega/T}+1} \int_0^\infty ds \, e^{-\tau(s)} n_{\rm dm}\big( r(x_o,s,\mu_s) \big)
\nonumber\\
&\approx&
 \frac{g_e \sigma_{\rm abs}}{e^{\omega/T}+1} \int_0^\lambda ds \, n_{\rm dm}\big( r(x_o,s,\mu_s) \big),
 \label{fe}
\eea
where the coordinate is defined in Figure~\ref{fig:schematic}, $\mu_s = \hat{s}\cdot \hat{x}_o$, and the optical depth is defined as
\bea
\tau(s) \equiv  \frac{1}{v} \int^s_0 ds' \, \alpha(s') .
\eea
In the second line of Eq.~\eqref{fe}, we have approximated the electron distribution by introducing a cut-off of the integration $\lambda$, defined as $\tau(\lambda) =1$. This is the mean free path of electron. 

The streaming velocity (or bulk velocity) of electron can be defined as 
\bea
\vec{v}_{\rm str} = \frac{\int \frac{d^3p}{(2\pi)^3} \vec{v} f_e}{\int \frac{d^3p}{(2\pi)^3}  f_e}.
\eea
In the limit where $\lambda < x_o$, an analytic expression for the streaming velocity can be obtained by using the approximate form of electron distribution,
\bea
\vec{v}_{\rm str} \simeq 
- \frac{\hat{x}_o}{\langle v^{-1} \rangle} \frac{\lambda}{6} \frac{\partial \ln n_{\rm dm}}{\partial x_o} \simeq \hat{x}_o \frac{\lambda}{6r_s} \langle v\rangle,
\eea
where $\langle v\rangle$ is the average velocity of electrons from Hawking radiation.
For the second expression, we have approximated $1/\langle v^{-1} \rangle \simeq \langle v \rangle$ and used Burkert profile with  $x_o < r_s$. 

The electron phase space distribution can be written in terms of the streaming velocity. 
Under the same approximation, $\langle v \rangle \simeq 1 / \langle v^{-1} \rangle$ and $\lambda < x_o < r_s$, we find
\bea
f_e(\vec{x}_o , p)  = f_0(x_o, p) \left[ 1 + \frac{3 v_{\rm str}}{\langle v \rangle} \mu_v \right]
\eea
where $\mu_v = \hat{v}\cdot \hat{x}_o = - \mu_s$, and 
\bea
f_0 (x_o,p) = \frac{g_e \sigma_{\rm abs}}{e^{\omega / T} +1 } \big[n_{\rm dm}(x_o) \lambda \big]. 
\eea

\section{Growth rate of hydromagnetic waves}\label{app:growth}
In this appendix, we compute the growth rate of hydromagnetic waves due to streaming relativistic electrons and positrons. 
We refer the interested reader to the textbook by~\citet{1986islp.book.....M} for detailed discussions. 

The Boltzmann equation for the electromagnetic excitation $M$ can be written as
\bea
\frac{df_M( \vec k)}{dt} 
= \alpha_M (\vec{k}) - \gamma_M(\vec{k}) f_M(\vec{k})
\eea
where each quantity is defined as
\bea
\alpha_M(\vec{k}) &=& + \sum_{s= - \infty}^\infty \int \frac{d^3p}{(2\pi)^3} \omega_M(\vec{p} ,\vec{k},s) f_{e}(\vec{k}+\vec{p}),
\\
\gamma_M(\vec{k}) &=& - \sum_{s = -\infty}^\infty \int \frac{d^3p}{(2\pi)^3} \omega_M(\vec{p} ,\vec{k},s) \hat{D}  f_{e}(\vec{p}).
\label{dec}
\eea
Here $\omega_M(\vec{p}, \vec{k},s)$ is the emission probability of the wave with a wave vector $\vec{k}$ at $s$-th harmonics.
The differential operator $\hat{D}$ is defined as
\bea
\hat{D} = \frac{\omega}{v} \frac{\partial}{\partial p} - \frac{\omega \cos\alpha - k_{||} v}{pv} \frac{\partial}{\partial \mu_\alpha}
\eea
where $\mu_\alpha = \cos\alpha = \hat{p} \cdot \hat{B}_0$, i.e an angle between the momentum of particle and the external magnetic field.
The mode $M$ grows when the damping rate $\gamma_M$ takes a negative value. 

For \alf wave, the transition probability is given as
\bea
\omega_A (\bfp,\bfk ,s) = 
\frac{\pi q^2 v_A}{4k |\cos\theta|} 
v_\perp^2 \delta(\omega - k_{||} v_{||} - s\Omega_e / \gamma ) , 
\eea
where $\cos\theta = \hat{k} \cdot \hat{B}_0$ is the angle between the wave vector and the external magnetic field, and $v_\perp$ is the velocity of particle perpendicular to the external magnetic field. 
From Eq.~\eqref{dec}, it is straightforward to find
\bea
\gamma_{A}(\vec{k})
&=& - \frac{\alpha_{\rm em} v_A^2}{4|k_{||}|} \int^\infty_{p_{\rm min}} d p \, p 
\Big(1 - \frac{p_{\rm min}^2}{p^2} \Big) 
\nonumber\\
&&\times
\left[
\left\{ 
p \frac{\partial}{\partial p}
- \mu_\alpha \frac{\partial}{\partial \mu_\alpha} 
\right\} f_{\rm sym}
+ \frac{v}{v_A} \frac{c_\theta}{|c_\theta|} \frac{\partial}{\partial \mu_\alpha} f_{\rm anti} 
\right]_{\mu_\alpha = \mu_R}
\label{alfven_streaming_pitch}
\eea
where  $\mu_R = - m_e \Omega_e / (p k_{||})$, $p_{\rm min} = m_e \Omega_e / |k_{||}|$, and we have defined
\bea
f_{\rm sym} &=& f(p,\mu_\alpha) + f(p, - \mu_\alpha), 
\\
f_{\rm anti} &=& f(p,\mu_\alpha) - f(p, - \mu_\alpha).
\eea
Note that the dispersion relation for the \alf wave is $\omega = v_A |k_{||}| < \Omega_i$ where $\Omega_i$ is the ion gyro-frequency. 
The damping rate for the other low frequency mode, magneto-acoustic mode, can be similarly obtained, and the result is the same as Eq.~\eqref{alfven_streaming_pitch}, except for the absence of $|\cos\theta|$ in the squared parenthesis. 

Now, we consider the approximate form of electron/positron phase space distribution in Leo T, $f_e(\vec{x}_o,p) = f_0 ( 1 +  3\frac{v_{\rm str} }{\langle v \rangle} \mu_v)$. 
For simplicity, let us consider the distribution function at a spatial position $\vec{x}_o$ such that the magnetic field coincides with the radial direction. In this case, $\mu_\alpha = \mu_v$ and the growth rate can be written as
\bea
\gamma_A (\vec{k}) 
\approx - \frac{\alpha_{\rm em} v_A^2}{2 |k_{||}|} \int^\infty_{0} d p \, p 
\left[
3 \frac{v_{\rm str}}{v_A} \frac{v}{\langle v\rangle} \frac{c_\theta}{|c_\theta|} 
-2
\right] f_0
\label{alfven_streaming_pitch}
\eea
where we have set $p_{\rm min} \to 0$ as typical momentum for the parameter space of our interest is larger than $p_{\rm min}$. 
From this expression, it is clear that the growth is only possible for $v_{\rm str} \gtrsim 2/3 v_A$.


\bibliographystyle{mnras}
\bibliography{ref}

\bsp	
\label{lastpage}
\end{document}